\documentclass[conference,a4paper]{APSIPA2021}
\usepackage{multirow}
\usepackage{amsmath}
\usepackage[psamsfonts]{amssymb}
\usepackage{amsxtra}
\usepackage{threeparttable}
\usepackage{multirow}
\usepackage{adjustbox}
\usepackage{graphicx}

\newcommand\blfootnote[1]{%
  \begingroup
  \renewcommand\thefootnote{}\footnote{#1}%
  \addtocounter{footnote}{-1}%
  \endgroup
}

\begin{document}

\title{I2CR: Improving Noise Robustness on Keyword Spotting using Inter-Intra Contrastive Regularization}

\author{%
\authorblockN{%
Dianwen Ng$^{1,2}$,
Jia Qi Yip$^{1,2}$,
Tanmay Surana$^{2}$,
Zhao Yang$^{3}$,
Chong Zhang$^{1}$,
Yukun Ma$^{1}$,\\
Chongjia Ni$^{1}$,
Eng Siong Chng$^{\dagger, 2}$ and 
Bin Ma$^{\dagger, 1}$
}
\authorblockA{%
$^1$Alibaba Group, Singapore\\
$^2$School of Computer Science and Engineering, Nanyang Technological University, Singapore\\
$^3$Faculty of Electronic and Information Engineering, Xi’an Jiaotong University, China
}
%
}

\maketitle
\thispagestyle{empty}

\begin{abstract}
Noise robustness in keyword spotting remains a challenge as many models fail to overcome the heavy influence of noises, causing the deterioration of the quality of feature embeddings. We proposed a contrastive regularization method called Inter-Intra Contrastive Regularization (I2CR) to improve the feature representations by guiding the model to learn the fundamental speech information specific to the cluster. This involves maximizing the similarity across Intra and Inter samples of the same class. As a result, it pulls the instances closer to more generalized representations that form more prominent clusters and reduces the adverse impact of noises. We show that our method provides consistent improvements in accuracy over different backbone model architectures under different noise environments. We also demonstrate that our proposed framework has improved the accuracy of unseen out-of-domain noises and unseen variant noise SNRs. This indicates the significance of our work with the overall refinement in noise robustness.
\end{abstract}

\section{Introduction}
\blfootnote{$\dagger$ These two authors contributed equally}

Keyword spotting (KWS) refers to the detection of keywords in utterances. KWS systems are ubiquitous with modern automated personal assistants like Siri, Google Assistant, Amazon Alexa, etc. These personal assistants are activated hands-free using keywords like “Hey Siri”, “Ok Google”, and “Alexa”. In recent years, there have been numerous successful works \cite{berg2021keyword, gong2021ast, majumdar2020matchboxnet, ng2022convmixer, ng2022small} on developing highly accurate KWS models. However, these works focus largely on clean or close-talking audio sets and often overlook noisy environments (i.e. below 0dB noises). As such, not many models in the existing literature are found to be robust under such noisy conditions. However, it is important that we understand the problem of noise robustness in keyword spotting as it is becoming prevalent for such applications to be used in noisy environments; for example, deploying the application in the middle of traffic (i.e. KWS is likely to be influenced by traffic noise). 

\begin{figure}[!htbp]
    \centering
    \includegraphics[width=87mm]{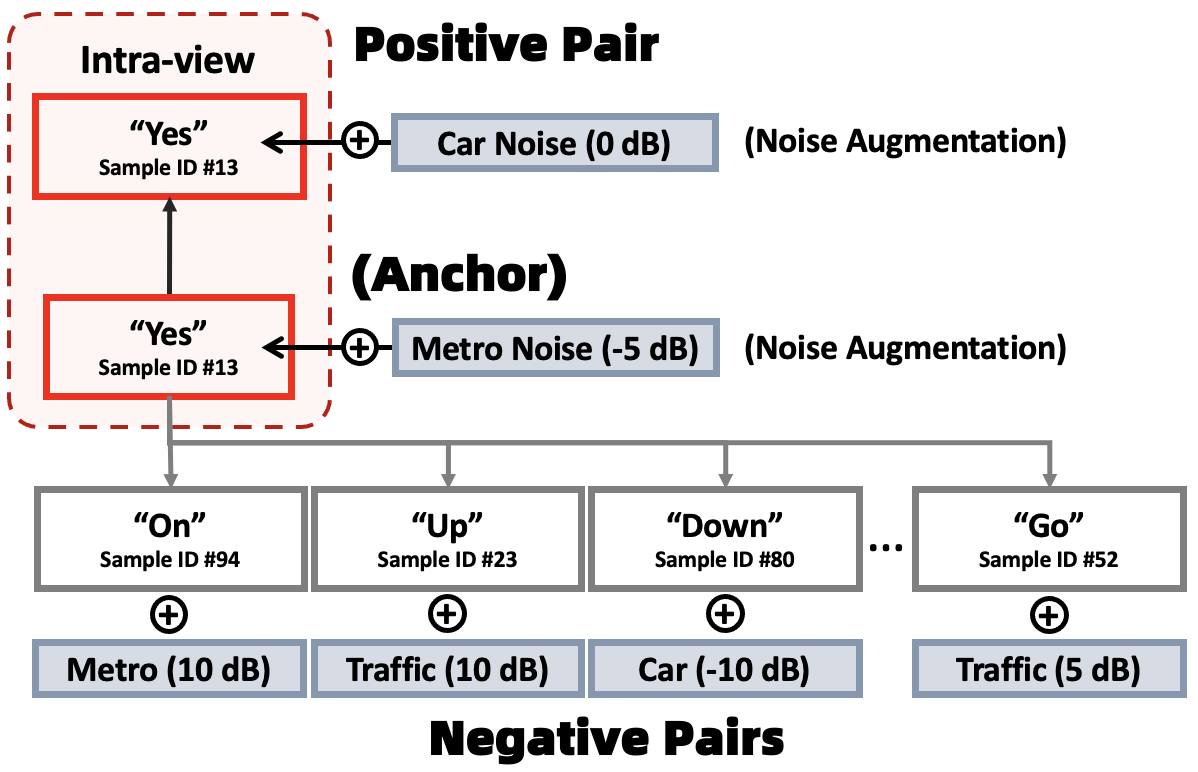}
    \caption{Graphical framework of conventional \textit{``N-pair"} self-supervised contrastive learning pairing view.}
    \label{fig:img0}
\end{figure}

A conventional approach to building a noise-robust KWS model involves conditioning the model to multiple levels of signal-to-noise ratio (SNR) noises (i.e. multi-conditioning training). However, without any additional computations for noise enhancement, the feature representations are still heavily influenced by the low SNR noise. As a result, multi-conditioning training depends hugely on the networks to adapt and generalize themselves to the selected SNR noise environments. The performance for different noises with variant SNRs outside the conditioning range is expected to be inferior, especially for the lower complexity KWS model that requires a smaller model footprint. To achieve better representations, the model needs to be more aware of the key representative features for the predicting classes and reduce the influence of the noises. This makes contrastive learning frameworks a potential means to improve model noise robustness. Typically, contrastive learning aims to extract the intrinsic features between two differently augmented views (called positives) of a sample by bringing the representations of the samples closer in embedding space while pushing away other samples (called negatives) \cite{chen2020simple, NEURIPS2020_f3ada80d}. A graphical understanding of the sampling framework is shown in Fig. \ref{fig:img0}. In short, each anchor has a positive sample with a different augmenting view of itself, and we can randomly pull one (triplet loss) or many negative samples (N-pair loss) of other non-related instances, usually taken from the mini-batch or in a memory queue. Then, this could conceptually reduce the noise influence to achieve higher-quality features by contrasting the positive pair with two differently augmented noises. The model is encouraged to learn the robust speech representations (i.e. keyword features) and repel the noise information in the input sample.

In fact, this idea has been utilized by the automatic speech recognition (ASR) community to learn better noise robust representations for their system. \cite{huang2022spiral, 9746929} builds a self-supervised contrastive architecture that employs contrastive learning with varying augmented noises in the pretraining to encourage the model to learn contextualized representations that are robust to noise. After which, fine-tuning is done on the labelled dataset. However, this strategy (i.e. pretraining followed by finetuning on a labelled dataset) may not be suitable in our case as we do not have other large external datasets for pretraining. As such, it will not be as meaningful as \cite{huang2022spiral, 9746929}. Therefore we do not consider this two-stage training in our work. 
On the other hand, using a contrastive regularizer with Cross Entropy (CE) helps to optimize our networks directly with less hassle and provides the additional motivation of better noise-robust representations.
\cite{Motiian2017UnifiedDS, dou2019domain, kim2021selfreg} use the contrastive loss as a regularizer to achieve domain generalization for image classification, with the general principle being that the additional constraint of bringing similar samples across domains closer together in embedding space leads to better domain generalization. Moreover, in the semi-supervised setting, pseudo-label-based contrastive regularization was also shown to outperform consistency regularization for image classification \cite{Lee_2022_CVPR}. \cite{al2021clar} uses the self-supervised contrastive loss in combination with CE for KWS (in the clean setting), leading to better performance than vanilla CE and faster in training.



In this work, we propose a contrastive regularization method, i.e. Inter-Intra Contrastive Regularization (I2CR), to improve the noise robustness of KWS models. The regularization method uses both augmentation within the given sample (i.e. intra-view) and other samples with the same label (i.e. inter-view) as positive pairs. Our method mitigates the problems with conventional cross-entropy and contrastive learning while encouraging better speech representations and faster training. Moreover, our method is simple to implement. Using additive noises along with other augmentations, the experimental results show that the proposed I2CR method leads to improved noise robustness to both in-domain and out-of-domain noises. Per our knowledge, this is the first work to tackle noise robustness in keyword spotting with contrastive learning. Our contributions are as follows:

\begin{enumerate}
\item We propose a Inter-Intra view Contrastive Regularization (I2CR) method to improve the noise robustness of KWS models.
\item We show that our method provides consistent improvements in accuracy over different backbone model architectures under different noise environments.
\item The experimental results also show that the proposed method consistently improves the accuracy for unseen out-of-domain noises with variant noise SNRs.
\end{enumerate}

\section{Methodology} \label{sec:method}
The conventional supervised learning model architecture focuses on identifying associative patterns to distinguish the data from multiple classes \cite{jaiswal2020invariant}. There is no additional constraint on the latent representations to prevent the model from picking up irrelevant factors such as gender, speaking style and background noise. The degeneration in the performance for noisy keyword spotting could be due to the failure to suppress such nuisance features that negatively impact the prediction. That is, the embedding of the latent representations could dominate heavily by the background noise. On the contrary, contrastive learning constraint itself to maximize information over the positive views and minimize the negative views. Through noise augmentation, we could create a dual-view from a sample and use contrastive learning constrain to encourage the model to focus more on extracting a generalized class features from two different noise augmented audios. This helps to reduce the amount of noise encoded in its latent features, making it robust to noise. Recent works exploited this to propose the self-supervised contrastive learning framework that consists of two-stage training \cite{khosla2020supervised, chen2020simple, qian2022contentvec}. We could first perform self-supervised pre-training followed by supervised finetuning on labelled data. However, this often requires heavy training resources (e.g. large pre-training dataset), and optimization is challenging (e.g. dimensional collapse in contrastive self-supervised learning). In this work, we offset the limitations of the former by combining supervised learning with contrastive learning as a regularizer so as to stabilize the training from dimensional collapse with stronger guidance from the supervised loss and improve the robustness of the model from the contrastive loss for noisy environments.

\subsection{Inter-Intra Contrastive Objective}
\begin{figure}[!htbp]
    \centering
    \includegraphics[width=87mm]{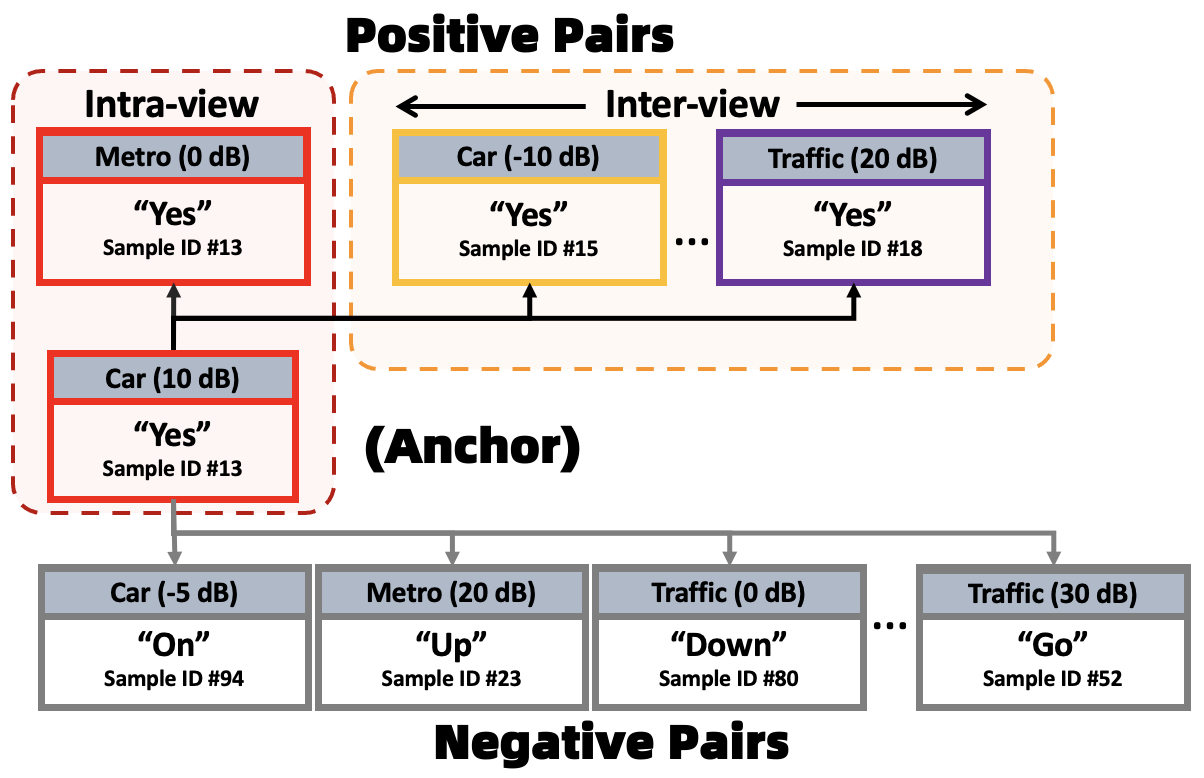}
    \caption{Graphical framework of our proposed I2CR anchor pairing view.}
    \label{fig:img1}
\end{figure}

Instead of using a single positive pair by augmenting the anchor with a different perturbating noise (i.e. intra-view contrasting) \cite{grill2020bootstrap}, we suggest pairing the anchor instance with several more positive views. An advantage of doing this allows the model to infer the embeddings of the class representations better with increasing references from the set of positive pairs \cite{aaai}. However, data augmentations of the same sample share a few other unfavourable factors, such as the speaker style, that could induce irrelevant biases to the embeddings. Conversely, our essential goal is to teach the model to encode the fundamental features that contain no other attributes except for the information of the class. As such, these positive pairs should be coming from different samples of distinct speakers from the same class group (i.e. inter-view contrasting). Since labels are provided during the supervised training, we can form the positive pairs (i.e. inter-intra view) from the minibatch by identifying samples of the same target label. An illustration of the graphical framework is shown in Fig. \ref{fig:img1}. Then, we compute the contrastive loss with 
$$ L{_{\text{I2CR}}} = \sum_{i\in I} \frac{-1}{\text{I}({P(i)})} \sum_{p \in P(i)} \log \frac{\exp(\text{sim}(z_i, z_p) / \tau)}{\sum_{n \notin P(i)}\exp(\text{sim}(z_i, z_n) / \tau)} $$ where $P(i)$ is the set of indices for all positive pairs, and we let sim($z_i, z_p$) be the dot product between normalize vector $z_i$ and $z_p$ (i.e. cosine similarity).

\subsection{Training Framework}
\begin{figure}[!htbp]
    \centering
    \includegraphics[width=87mm, height=69mm]{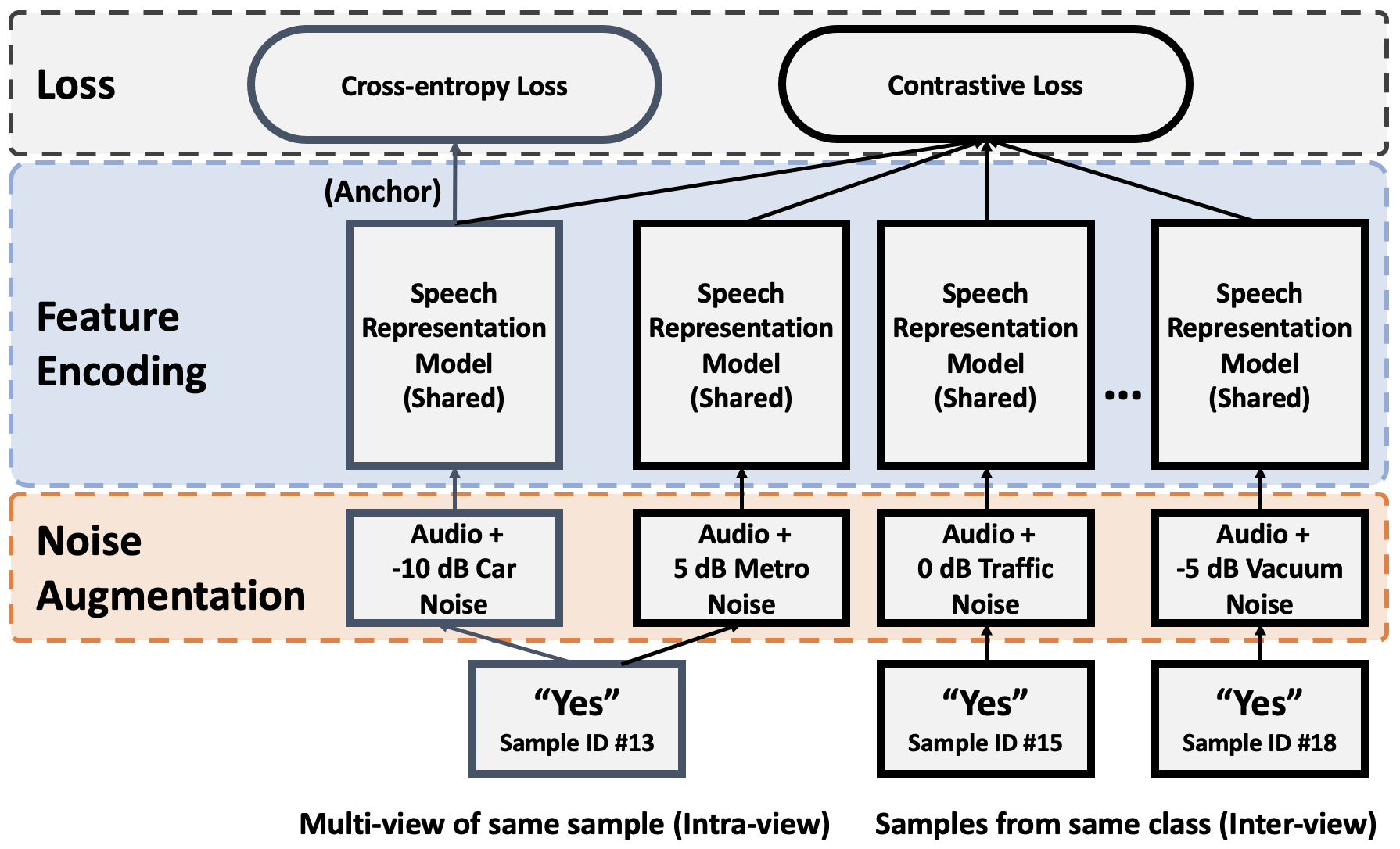}
    \caption{Framework of our proposed supervised model with I2CR regularizer.}
    \label{fig:img2}
\end{figure}
In general, our model is trained with the combine loss functions
$$L = L_{CE} + \alpha L_{I2CR}$$ where $L_{CE}$ is the cross-entropy loss for supervised learning and $L_{I2CR}$ is the contrastive loss as the regularizer. $\alpha$ is the weight given to the regularizer that weighs the degree of the additional contrastive constrain.  

To maintain consistency and allow for a fair comparison between the baseline (i.e. vanilla supervised learning) and the model trained with our proposed regularizer, we fixed all hyperparameters for all experiments. Here, we show our results with three widely used feature encoding models, namely EfficientNet-B0 \cite{tan2019efficientnet}, ResNet-18 \cite{he2016deep} and Keyword Transformer (KWT-1) \cite{berg2021keyword}. EfficientNet-B0 and ResNet-18 encoders are convolutional-based and are the state-of-the-arts (SOTA) models for a diverse range of tasks, including image and audio event classification. KWT-1 model is based on the recent trending transformer architecture designed for keyword spotting. An overview of our proposed framework is presented in Fig. \ref{fig:img2}.

To ensure meaningful and effective contrastive representation learning \cite{tian2020makes}, we carefully design our data augmentation steps as in the following list.
\begin{itemize}
  \item \textbf{Noise Augmentation}: randomly select segment of noise from our noise dataset (FSD50K) and adds it to our keyword audios. The noise is added with randomly determined SNR between (-10, 30).
  \item \textbf{Time-frequency Masking} \cite{park2019specaugment}: randomly select a small segment from the time-frequency audio and set the signal values of the masking block of frequency channels, and masking block of time steps to be zero. We fixed the number of maskings for time and frequency to be 2. The maximum size of masked segment for time is 25, and frequency is 7. 
  \item \textbf{Time Shifting}: randomly shift the audio samples forwards or backwards. Samples that roll beyond the last position are reintroduced to the first position. The degree and direction of the shifts were randomly choosen between (-100ms, 100ms) for each audio. 
  \item \textbf{Speed Perturbation} \cite{ko2015audio}: randomly modify the speed of a signal by resampling the signal. The speed function of Pytorch Sox was used for this and we modify the speed from 90\% to 110\% of the original rate.
\end{itemize}

The above-mentioned perturbation is performed to every augmenting view of the audio. However, we realize that the implementation of inter-view contrasting is extremely challenging in the early stage. This is due to the early initialization of random weights as we train the model from scratch, and the vector representations of the instance are consequently much more random. Imposing the constraint to maximize the similarity between two random vectors of a different (positive) sample, even though they come from the same class, is non-trivial. Hence, we set the degree of regularization in $\alpha$ to be zero for the first epoch. Subsequently, we increase $\alpha$ at a linear rate (i.e. current epoch / total number of epochs), and we clip the maximum degree to 0.5. This helps to stabilize our proposed framework.

\section{Experiments}
\begin{table*}[] \centering\small
\renewcommand{\arraystretch}{1.3}
\tabcolsep=0.16cm
\caption{Comparison of the performance in accuracy of different encoding models between the vanilla supervised learning model and the proposed I2CR framework with specific noise types and variant SNRs in the command (35 classes) test set.}
\begin{tabular}{l|cccccccccccc}
\hline
\multirow{3}{*}{Models (Method)} & \multicolumn{12}{c}{Accuracy under different SNRs (dB)}                                                              \\ \cline{2-13} 
                        & \multicolumn{4}{c|}{Traffic}           & \multicolumn{4}{c|}{Metro}             & \multicolumn{4}{c}{Car} \\ \cline{2-13} 
                        & -10 & -5 & 0 & \multicolumn{1}{c|}{20} & -10 & -5 & 0 & \multicolumn{1}{c|}{20} & -10   & -5   & 0  & 20  \\ \hline
EfficientNet-B0 (Base)         & 0.840   & 0.905  & 0.944 & \multicolumn{1}{c|}{0.966}  & 0.881   & 0.930  & 0.951 & \multicolumn{1}{c|}{0.968}  & 0.825     & 0.896    & 0.928  & 0.957   \\
EfficientNet-B0 (Intra Reg.)         & 0.858   & 0.918  & 0.948 & \multicolumn{1}{c|}{0.971}  & 0.895   & 0.939  & 0.958 & \multicolumn{1}{c|}{\textbf{0.973}}  & 0.841     & 0.905    & 0.937  & 0.960   \\
EfficientNet-B0 (I2CR Reg.)        & \textbf{0.867}   & \textbf{0.925}  & \textbf{0.950} & \multicolumn{1}{c|}{\textbf{0.972}}  & \textbf{0.903}   & \textbf{0.942}  & \textbf{0.960} & \multicolumn{1}{c|}{\textbf{0.973}}  & \textbf{0.850}     & \textbf{0.910}    & \textbf{0.940}  & \textbf{0.961}   \\ \hline
ResNet-18 (Base)              & 0.860   & 0.921  & 0.950 & \multicolumn{1}{c|}{0.971}  & 0.898   & 0.942  & 0.957 & \multicolumn{1}{c|}{0.974}  & 0.841     & 0.909    & 0.937  & 0.961   \\
ResNet-18 (Intra Reg.)              & 0.868   & 0.927  & 0.954 & \multicolumn{1}{c|}{\textbf{0.975}}  & 0.908   & 0.942  & 0.963 & \multicolumn{1}{c|}{\textbf{0.977}} & 0.853     & 0.911    & 0.942  & 0.965   \\
ResNet-18 (I2CR Reg.)              & \textbf{0.871}   & \textbf{0.931}  & \textbf{0.956} & \multicolumn{1}{c|}{0.974}  & \textbf{0.911}   & \textbf{0.947}  & \textbf{0.965} & \multicolumn{1}{c|}{\textbf{0.977}}  & \textbf{0.862}     & \textbf{0.916}    & \textbf{0.944}  & \textbf{0.966}   \\ \hline
KWS Transformer (Base)        & 0.757   & 0.860  & 0.909 & \multicolumn{1}{c|}{0.954}  & 0.807   & 0.891  & 0.926 & \multicolumn{1}{c|}{0.956}  & 0.745     & 0.848    & 0.897  & 0.941   \\
KWS Transformer (Intra Reg.)        & 0.766   & 0.867  & 0.912 & \multicolumn{1}{c|}{\textbf{0.956}}  & 0.822   & 0.900  & 0.928 & \multicolumn{1}{c|}{\textbf{0.959}}  & 0.757     & 0.851    & 0.901  & 0.945   \\
KWS Transformer (I2CR Reg.)        & \textbf{0.771}   & \textbf{0.869}  & \textbf{0.913} & \multicolumn{1}{c|}{0.955}  & \textbf{0.825}   & \textbf{0.904}  & \textbf{0.930} & \multicolumn{1}{c|}{0.958}  & \textbf{0.759}     & \textbf{0.855}    & \textbf{0.902}  & \textbf{0.946}   \\ \hline

\end{tabular}
\label{tbl:1}
\end{table*}

\begin{table*}[] \centering\small
\renewcommand{\arraystretch}{1.3}
\tabcolsep=0.16cm
\caption{Comparison of the performance in accuracy of different encoding models between the vanilla supervised learning model and the proposed IC2R framework with specific noise types and variant SNRs in the command (10 classes) test set.}
\begin{tabular}{l|cccccccccccc}
\hline

\multirow{3}{*}{Models (Method)} & \multicolumn{12}{c}{Accuracy under different SNRs (dB)}                                                              \\ \cline{2-13} 
                        & \multicolumn{4}{c|}{Traffic}           & \multicolumn{4}{c|}{Metro}             & \multicolumn{4}{c}{Car} \\ \cline{2-13} 
                        & -10 & -5 & 0 & \multicolumn{1}{c|}{20} & -10 & -5 & 0 & \multicolumn{1}{c|}{20} & -10   & -5   & 0  & 20  \\ \hline
EfficientNet-B0 (Base)         & 0.870   & 0.928  & 0.953 & \multicolumn{1}{c|}{0.976}  & 0.900   & 0.941  & 0.958 & \multicolumn{1}{c|}{0.978}  & 0.848     & 0.908    & 0.934  & 0.961   \\
EfficientNet-B0 (Intra Reg.)         & 0.893   & 0.937  & 0.960 & \multicolumn{1}{c|}{\textbf{0.981}}  & 0.918   & 0.950  & 0.966 & \multicolumn{1}{c|}{\textbf{0.982}}  & 0.867     & 0.922    & 0.947  & 0.969   \\
EfficientNet-B0 (I2CR Reg.)        & \textbf{0.897}   & \textbf{0.940}  & \textbf{0.962} & \multicolumn{1}{c|}{\textbf{0.981}}  & \textbf{0.924}   & \textbf{0.953}  & \textbf{0.971} & \multicolumn{1}{c|}{0.981}  & \textbf{0.878}     & \textbf{0.924}    & \textbf{0.949}  & \textbf{0.971}   \\ \hline
ResNet-18 (Base)              & 0.887   & 0.942  & 0.960 & \multicolumn{1}{c|}{0.978}  & 0.922   & 0.956  & 0.970 & \multicolumn{1}{c|}{0.982}  & 0.872     & 0.922    & 0.953  & 0.968   \\
ResNet-18 (Intra Reg.)              & 0.891   & 0.950  & 0.964 & \multicolumn{1}{c|}{\textbf{0.984}}  & 0.925   & 0.959  & 0.972 & \multicolumn{1}{c|}{\textbf{0.985}}  & 0.895     & 0.937    & 0.954  & \textbf{0.975}   \\
ResNet-18 (I2CR Reg.)              & \textbf{0.901}   & \textbf{0.953}  & \textbf{0.968} & \multicolumn{1}{c|}{0.983}  & \textbf{0.932}   & \textbf{0.962}  & \textbf{0.975} & \multicolumn{1}{c|}{\textbf{0.985}}  & \textbf{0.904}     & \textbf{0.941}    & \textbf{0.958}  & 0.974   \\ \hline
KWS Transformer (Base)        & 0.765   & 0.866  & 0.916 & \multicolumn{1}{c|}{0.957}  & 0.806   & 0.877  & 0.927 & \multicolumn{1}{c|}{0.959}  & 0.743     & 0.840    & 0.896  & 0.944   \\
KWS Transformer (Intra Reg.)        & 0.793   & 0.887  & 0.926 & \multicolumn{1}{c|}{\textbf{0.966}}  & 0.826   & 0.895  & 0.933 & \multicolumn{1}{c|}{\textbf{0.971}}  & 0.771     & 0.851    & 0.910  & 0.958   \\
KWS Transformer (I2CR Reg.)        & \textbf{0.805}   & \textbf{0.891}  & \textbf{0.929} & \multicolumn{1}{c|}{\textbf{0.966}}  & \textbf{0.837}   & \textbf{0.903}  & \textbf{0.939} & \multicolumn{1}{c|}{0.970}  & \textbf{0.782}     & \textbf{0.860}    &  \textbf{0.916}  & \textbf{0.961}   \\ \hline


\end{tabular}
\label{tbl:2}
\end{table*}
\subsection{Data Description}
We build our work based on the two commonly used subsets (i.e. 35 classes and 10 classes) from Google Speech Command dataset V2 \cite{warden2018speech}, which consists of 1-second long keyword utterances sampled at 16kHz. Furthermore, there are overall 105,000 samples of 35 different words. The 10 classes subset is made up of ``up", ``down", ``left", ``right", ``yes", ``no", ``on", ``off", ``go", and ``stop", and the 35 classes dataset contains all of the available words. We use the same 80/10/10 (train/validation/test) split as \cite{warden2018speech, zhang2017hello} for better comparison and easy work reproduction. As our intention is to find out if our regularizer helps improve the noise robustness of the keyword spotting model in general, we are interested in evaluating the performance with noises that are not used during the training phase (i.e. out-of-domain noises). This is strongly motivated by real-world applications as it is impossible to see all noises during training. Therefore, we keep the training noise environment of our experiments to focus mainly on three common outdoor scenarios, i.e. traffic, metro and car, and SNRs varying from -10dB to 30dB. Then in the later section, we assess the models with noises not only from the former three but also with other domain noises, and ranges of SNRs that are not adapted during training. We collect our noise dataset from FSD50K (Freesound) \cite{fonseca2021fsd50k}, an open dataset of human-labelled sound events. Altogether, we extracted 905 instances from the official train, and 487 samples from the official evaluation set for testing. In addition, we use the background noises provided by the Google Speech Commands as our out-of-domain noises. 

\subsection{Implementation Details}
Our model uses a 64-dimensional log Mel filterbanks (FBanks) as input derived from a 25ms window size kernel and a 10ms shift. We fixed the resolution of our FBanks to 1 second, which is equivalent to the size of $98 \times 64$. For utterances that are shorter than 1 second will be zero-padded to the right, while longer ones will be cut and trimmed. During training, data augmentation is performed as described in section \ref{sec:method}. All models are trained with a batch size of 128, and we set the initial learning rate to be 5e-4. We decay the learning rate with cosine annealing, and the minimum is pre-determined to be 1e-12. Adam optimizer is used in the optimization process. Lastly, we trained our models for 100 epochs to ensure the convergence of all experiments.

\begin{figure*}[!hbt]
    \centering
    \includegraphics[width=165mm,height=68mm]{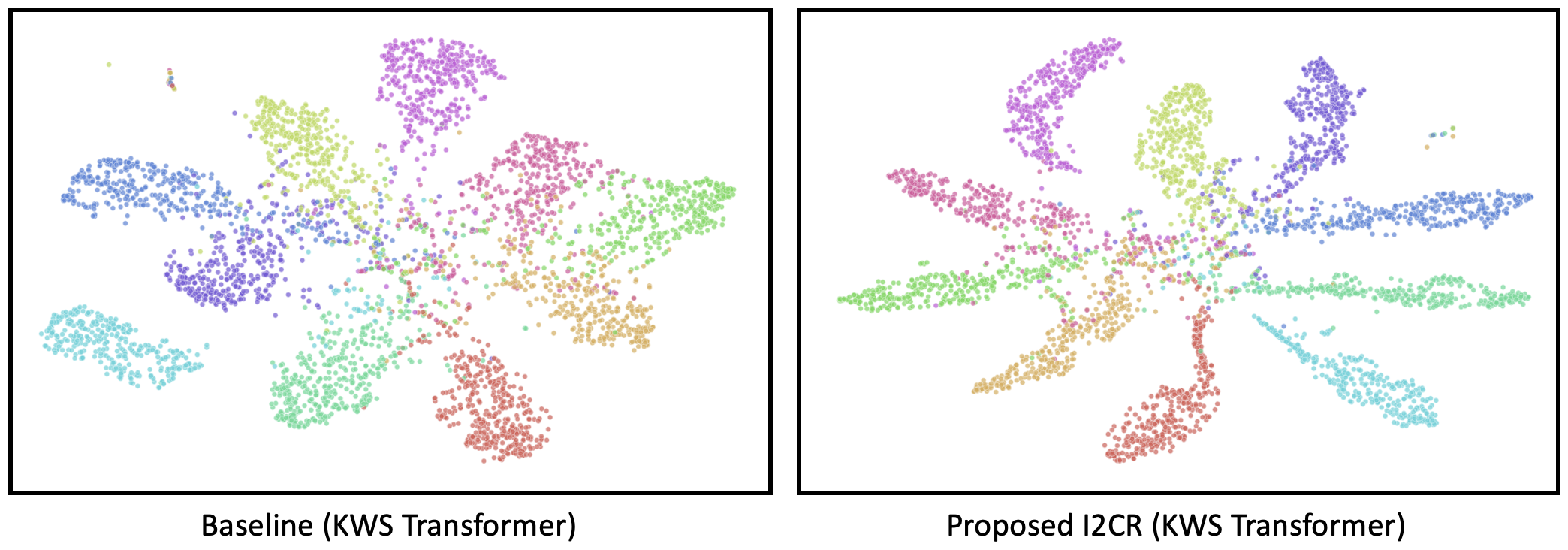}
    \caption{t-SNE plot of the latent representations from keyword transformer and noisy command of the containing 10 keyword classes (0 dB) test set}
    \label{fig:img3}
\end{figure*}

\begin{table*}[!htb]\centering\small
\renewcommand{\arraystretch}{1.3}
\tabcolsep=0.38cm
\caption{Comparison of the performance in accuracy with the noisy audio test set evaluated outside of the range of training SNRs}
\begin{tabular}{l|cccccccc}
\hline
\multirow{3}{*}{Models (No. of Classes)} &
  \multicolumn{8}{c}{Accuracy under different SNRs (dB)} \\ \cline{2-9} 
 &
  \multicolumn{2}{c|}{-30} &
  \multicolumn{2}{c|}{-20} &
  \multicolumn{2}{c|}{-15} &
  \multicolumn{2}{c}{Clean} \\ \cline{2-9} 
 &
  \multicolumn{1}{c|}{Base} &
  \multicolumn{1}{c|}{I2CR} &
  \multicolumn{1}{c|}{Base} &
  \multicolumn{1}{c|}{I2CR} &
  \multicolumn{1}{c|}{Base} &
  \multicolumn{1}{c|}{I2CR} &
  \multicolumn{1}{c|}{Base} &
  I2CR \\ \hline
EfficientNet-B0 (35) &
  \multicolumn{1}{c|}{0.276} &
  \multicolumn{1}{c|}{\textbf{0.323}} &
  \multicolumn{1}{c|}{0.567} &
  \multicolumn{1}{c|}{\textbf{0.618}} &
  \multicolumn{1}{c|}{0.711} &
  \multicolumn{1}{c|}{\textbf{0.763}} &
  \multicolumn{1}{c|}{0.969} &
  \textbf{0.974} \\
ResNet-18 (35) &
  \multicolumn{1}{c|}{0.316} &
  \multicolumn{1}{c|}{\textbf{0.326}} &
  \multicolumn{1}{c|}{0.599} &
  \multicolumn{1}{c|}{\textbf{0.631}} &
  \multicolumn{1}{c|}{0.745} &
  \multicolumn{1}{c|}{\textbf{0.767}} &
  \multicolumn{1}{c|}{0.974} &
  \textbf{0.977} \\
KWS Transformer (35) &
  \multicolumn{1}{c|}{0.196} &
  \multicolumn{1}{c|}{\textbf{0.211}} &
  \multicolumn{1}{c|}{0.449} &
  \multicolumn{1}{c|}{\textbf{0.473}} &
  \multicolumn{1}{c|}{0.608} &
  \multicolumn{1}{c|}{\textbf{0.637}} &
  \multicolumn{1}{c|}{0.950} &
  \textbf{0.955} \\ \hline
EfficientNet-B0 (10) &
  \multicolumn{1}{c|}{0.355} &
  \multicolumn{1}{c|}{\textbf{0.411}} &
  \multicolumn{1}{c|}{0.623} &
  \multicolumn{1}{c|}{\textbf{0.681}} &
  \multicolumn{1}{c|}{0.754} &
  \multicolumn{1}{c|}{\textbf{0.806}} &
  \multicolumn{1}{c|}{0.976} &
  \textbf{0.983} \\
ResNet-18 (10) &
  \multicolumn{1}{c|}{0.378} &
  \multicolumn{1}{c|}{\textbf{0.407}} &
  \multicolumn{1}{c|}{0.664} &
  \multicolumn{1}{c|}{\textbf{0.690}} &
  \multicolumn{1}{c|}{0.784} &
  \multicolumn{1}{c|}{\textbf{0.814}} &
  \multicolumn{1}{c|}{0.982} &
  \textbf{0.987} \\
KWS Transformer (10) &
  \multicolumn{1}{c|}{0.248} &
  \multicolumn{1}{c|}{\textbf{0.285}} &
  \multicolumn{1}{c|}{0.469} &
  \multicolumn{1}{c|}{\textbf{0.539}} &
  \multicolumn{1}{c|}{0.618} &
  \multicolumn{1}{c|}{\textbf{0.673}} &
  \multicolumn{1}{c|}{0.945} &
  \textbf{0.968} \\ \hline
\end{tabular}
\label{tbl:3}
\end{table*}

\begin{table*}[!htb]\centering\small
\renewcommand{\arraystretch}{1.3}
\tabcolsep=0.38cm
\caption{Comparison of the performance in accuracy with unseen out-of-domain command background noises}
\begin{tabular}{l|cccccccc}
\hline
\multirow{3}{*}{Models (No. of Classes)} &
  \multicolumn{8}{c}{Accuracy under different SNRs (dB)} \\ \cline{2-9} 
 &
  \multicolumn{2}{c|}{-10} &
  \multicolumn{2}{c|}{-5} &
  \multicolumn{2}{c|}{0} &
  \multicolumn{2}{c}{20} \\ \cline{2-9} 
 &
  \multicolumn{1}{c|}{Base} &
  \multicolumn{1}{c|}{I2CR} &
  \multicolumn{1}{c|}{Base} &
  \multicolumn{1}{c|}{I2CR} &
  \multicolumn{1}{c|}{Base} &
  \multicolumn{1}{c|}{I2CR} &
  \multicolumn{1}{c|}{Base} &
  I2CR \\ \hline
EfficientNet-B0 (35) &
  \multicolumn{1}{c|}{0.618} &
  \multicolumn{1}{c|}{\textbf{0.664}} &
  \multicolumn{1}{c|}{0.787} &
  \multicolumn{1}{c|}{\textbf{0.821}} &
  \multicolumn{1}{c|}{0.876} &
  \multicolumn{1}{c|}{\textbf{0.894}} &
  \multicolumn{1}{c|}{0.957} &
  \textbf{0.966} \\
ResNet-18 (35) &
  \multicolumn{1}{c|}{0.645} &
  \multicolumn{1}{c|}{\textbf{0.676}} &
  \multicolumn{1}{c|}{0.813} &
  \multicolumn{1}{c|}{\textbf{0.831}} &
  \multicolumn{1}{c|}{0.889} &
  \multicolumn{1}{c|}{\textbf{0.903}} &
  \multicolumn{1}{c|}{0.966} &
  \textbf{0.970} \\
KWS Transformer (35) &
  \multicolumn{1}{c|}{0.458} &
  \multicolumn{1}{c|}{\textbf{0.490}} &
  \multicolumn{1}{c|}{0.689} &
  \multicolumn{1}{c|}{\textbf{0.701}} &
  \multicolumn{1}{c|}{0.816} &
  \multicolumn{1}{c|}{\textbf{0.824}} &
  \multicolumn{1}{c|}{0.940} &
  \textbf{0.945} \\ \hline
EfficientNet-B0 (10) &
  \multicolumn{1}{c|}{0.686} &
  \multicolumn{1}{c|}{\textbf{0.727}} &
  \multicolumn{1}{c|}{0.826} &
  \multicolumn{1}{c|}{\textbf{0.858}} &
  \multicolumn{1}{c|}{0.892} &
  \multicolumn{1}{c|}{\textbf{0.920}} &
  \multicolumn{1}{c|}{0.967} &
  \textbf{0.976} \\
ResNet-18 (10) &
  \multicolumn{1}{c|}{0.725} &
  \multicolumn{1}{c|}{\textbf{0.761}} &
  \multicolumn{1}{c|}{0.854} &
  \multicolumn{1}{c|}{\textbf{0.886}} &
  \multicolumn{1}{c|}{0.917} &
  \multicolumn{1}{c|}{\textbf{0.931}} &
  \multicolumn{1}{c|}{0.971} &
  \textbf{0.979} \\
KWS Transformer (10) &
  \multicolumn{1}{c|}{0.482} &
  \multicolumn{1}{c|}{\textbf{0.577}} &
  \multicolumn{1}{c|}{0.672} &
  \multicolumn{1}{c|}{\textbf{0.742}} &
  \multicolumn{1}{c|}{0.788} &
  \multicolumn{1}{c|}{\textbf{0.840}} &
  \multicolumn{1}{c|}{0.928} &
  \textbf{0.941} \\ \hline
\end{tabular}
\label{tbl:4}
\end{table*}

\subsection{Experimental Results}
In general, all models have well converged with the best accuracy attained for the 10-classes models to be as high as about 98\%, and 35-classes models to be around 97\% under the 20 dB noise environment. We observed that ResNet-18 consistently performs the best over the two subsets even with the baseline architectural setup with at least 84\% accuracy for the 35-classes and 87\% accuracy for the 10-classes on the lowest -10dB noise. This is expected since it has the largest model size among the three. Our keyword transformer with the smallest model footprint slightly underperforms the other two on the lowest decibel noise of -10dB with a minimally 74\% accuracy over the two subsets. Regardless, this is still considerably well taking into account that it is under such an intense low level SNR noise environment. However, we observed an overall improvement in performance as we added our proposed contrastive regularizer. The enhanced performance is consistent over the two datasets. Although there is a smaller boost in accuracy for cleaner audio (i.e. 20dB noise), the gain from the extra constraint is most significant under lower SNR noises. Furthermore, we note that the improvement is as much as 4\% absolute from the keyword transformer for the -10dB car noise in the 10 classes set. These improvements are also well noticeable for other models of other noises. Moreover, while the middle row that implements intra-view contrastive (i.e. augmented view of the same sample) helps to boost the accuracy over lower SNR noise, I2CR appears superior and outperforms all. Besides, we attempt to visually understand the performance gain in our work with a graphical plot of the latent representations from the keyword transformer as seen in Fig. \ref{fig:img3}. The data samples used are based on the command test set with augmented 0dB noises from our FSD50K evaluation set. We took the latent representations at the bottleneck of our classifier before softmax operation and project it into two dimensional space using t-SNE algorithm \cite{JMLR:v9:vandermaaten08a} to complete the plot. From Fig. \ref{fig:img3}, we recognised that the clusters in I2CR are more prominent and well defined than the baseline. There are fewer overlappings, with more concentrated sample points found within clusters in the proposed I2CR. This indicates that our I2CR has richer embeddings of class representations that are less susceptible to nuisance noise. As a whole, we conclude that the contrastive regularizer that maximizes the agreement over the noise augmented positive pairs improves the latent speech representations by reducing the tendency of nuisance factors. The outcome could be more prominent when we contrast it with different samples of the same class. In addition, all these are achieved without harming the original performance and any modification to the model architecture.

\section{Post-Methodology Studies}
In this section, we address two striking and vital questions to show the significance of our work in improving the noise robustness as a whole. 
\begin{enumerate}
\item  Does the model also improves its robustness to other SNR noises beyond the conditioning range?
\item Does the model achieve better noise robustness to out-of-domain noises?
\end{enumerate}

\subsection{Robustness to SNR range beyond the training set}
\begin{table}[!ht]\centering
\renewcommand{\arraystretch}{1.3}
\tabcolsep=0.35cm
\caption{Comparison of the performance in accuracy with other architectures on clean Google Speech Commands (35 commands). These architectures are from the existing works that train and test their models on the clean dataset, whereas our models have never seen the clean domain environment.}
\begin{tabular}{l|c}
\hline
Method(s)                        & \begin{tabular}[c]{@{}c@{}}Test Accuracy\\ (Clean)\end{tabular} \\ \hline
TDNN (+ $x$vector)-SpeechBrain \cite{ravanelli2021speechbrain}             & 0.974                                                           \\
Conformer-ESPnet \cite{arora2022espnet}              & 0.975                                                           \\
Audio Spectrogram Transformer \cite{gong2021ast} & \textbf{0.981}                                                           \\
Branchformer \cite{peng2022branchformer}                 & 0.973                                                           \\
(Our) EfficientNet-B0-I2CR$^\star$          & 0.974 \\
(Our) ResNet-18-I2CR$^\star$          & \underline{0.977}                                                           \\ \hline
\end{tabular}
\label{table:clean2}
\end{table}

We tested our model based on the command test set with noises augmented on more extreme SNRs (i.e. -30dB, -20dB, -15dB). The more adverse condition aligns with the interest of the industry as it becomes exceptionally challenging to build a good keyword spotting model under such noise levels. From Table \ref{tbl:3}, we notice that the performance for all models has dropped drastically under -30dB. The degradation is worst on the 35 classes model. However, we saw that I2CR regularized model seems to be significantly better, with an absolute gain in accuracy of 4\% on average for the 10-classes subset, and 2.4\% gain for the 35-classes subset under -30dB noises. Besides, the gain becomes more obvious when it gets slightly less noisy and closer to the training SNR range. In addition, the proposed models are also better at clean audio, while it has never seen clean audio during training. This is further shown in Table \ref{table:clean2}, as we noticed that our model's performance on the clean audio is also comparable to the existing works that have trained and tested their models on the clean data. This could be the result of the constraint that forces the model to focus on encoding richer class representations of higher quality. 

\subsection{Robustness to out-of-domain noises}

Here, we tested our model based on the command test set with noises augmented from the background noises provided by the command data. The model has never seen this noise dataset during training. We show the results in Table \ref{tbl:4} with the SNRs similar to the training range. Likewise, we observed the performance drops with out-of-domain noises on all models. The results are anticipated since the models were never conditioned on the noises. However, models that are trained with the proposed I2CR regularization have achieved higher accuracy in comparison to the baseline. On average, the models of 10-classes improve by an absolute gain of 5.7\% under -10dB noises, and the improvements remain significant even as it becomes less noisy. Furthermore, we saw a similar trend in the models of 35-classes (i.e. with an absolute gain of 3.6\% under -10dB noises) as well.

\section{Conclusions}
In conclusion, we improve the noise robustness of the keyword spotting model trained using the vanilla supervised framework with our proposed Inter-Intra contrastive regularization. With extensive experiments conducted, we demonstrated a general trend of achieving higher accuracy with our suggested constraint that helps reduce the influence of nuisance noises. The framework also results in better noise robustness in comparison to the baseline with out-of-domain noises and broader SNR ranges, even when it was not conditioned during training. Besides, there is no additional complexity added to the original model architecture.

\section*{Acknowledgment}
This work was supported by Alibaba Group through Alibaba Innovative Research (AIR) Program and Alibaba-NTU Singapore Joint Research Institute (JRI), Nanyang Technological University, Singapore. Special thank Ruixi Zhang for the unwavering support.

\bibliographystyle{ieeetr} 
\bibliography{ref} 



\end{document}